# Cosmological Brightness Distribution Fits of Gamma Ray Burst Sources

I. Horváth[1,2], P. Mészáros [1] and A. Mészáros [3]

[1] Pennsylvania State University, 525 Davey Lab, University Park, PA 16803
[2] Central Research Inst. for Physics, PO Box 49, H-1525 Budapest, Hungary
[3] Dept. of Astronomy, Charles University, 150 00 Prague 5, Czech Republic

## ABSTRACT

We discuss detailed fits of the BATSE and PVO gamma-ray burst peak-flux distributions with Friedman models taking into account possible density evolution and standard candle or power law luminosity functions. A chi-square analysis is used to estimate the goodness of the fits and we derive the significance level of limits on the density evolution and luminosity function parameters. Cosmological models provide a good fit over a range of parameter space which is physically reasonable.

## 1. INTRODUCTION

Gamma ray burst sources are distributed with a very high level of isotropy (Fishman *et al.* , 1994), which is compatible with either a cosmological origin or an extended galactic halo origin. The brightness distribution is another indicator used to characterize the spatial distribution in distance, and this can be used to further test the distance scale hypotheses. This is generally done by investigating the functional behavior of the integral number $N$ of sources with peak photon flux rates $P$ above a certain value, $N(>P)$, or of the peak count rate divided by the threshold rate $N(>C_{max}/C_{min})$, or of the corresponding differential distributions. Comparisons of observed versus expected values in Friedman cosmologies have been discussed, e.g., by Mao & Paczyński (1992), Dermer (1992), Piran (1992) and Wasserman (1992). Statistical fits to a $\log N - \log P$ or $\log N - \log C$ distribution have been done by Loredo & Wasserman (1992), Wickramasinghe, *et al.* . (1993), Cohen & Piran (1994), Emslie & Horack (1994), Horack, Emslie & Hartmann (1995), Fenimore & Bloom (1995), Rutledge, *et al.* (1995). One of the main questions that such fits must address is the size of the parameter space region which is compatible with a cosmological distribution, and whether such parameters are reasonable. If the acceptable region contains physically plausible parameters and is not too restricted, one may assume the consistency of the observations with a general type of models; if on the other hand the acceptable region is very small and/or populated mainly by physically implausible parameters, fine-tuning would be required to fit the observations, and the case for consistency with those models is weaker. Such consistency, and absence of fine-tuning, is a requirement expected of any successful model of the GRB distribution, whether cosmological or galactic. Here, we shall address only the question of the consistency of the number distribution under the hypothesis of a cosmological distribution.

So far, most cosmological fits have been made with relatively specialized models, generally either with non-evolving or evolving density standard candle models, or with non-evolving luminosity functions. Limits on the luminosity function were investigated in cosmology with a pure density evolution by Horack, Emslie & Hartmann (1995) using a method of moments. In Euclidean space, limits have been investigated by Horack, Emslie & Meegan (1995), Ulmers & Wijers (1995) and Ulmer, Wijers & Fenimore (1995). Most cosmological calculations have used either the 1B or the 2B BATSE data base, and did not include the PVO information (see, however, Cohen & Piran, 1995, Fenimore & Bloom, 1995). In the present paper we make detailed chi-squared fits of the observed brightness distribution directly to specific models of the cosmological burst brightness distribution. We use both standard candle and power-law luminosity function models with a density evolving as a power law of the scale factor, for a wide range of density evolution exponents, luminosities and luminosity spreads, assuming either a brightness limited or redshift limited cases for various maximum redshifts for the source distribution. This is done both using the BATSE 2B catalogue of sources (Meegan, *et al.* , 1994), and combining the BATSE catalogue with information published for the PVO counts (Fenimore and Bloom, 1995). The significance levels of the various cosmological fits is discussed for both the 2B and the expanded burst sample.

## 2. COSMOLOGICAL DISTRIBUTION MODELS

Analytical expressions for the integral burst number counts $N(>P)$ with peak photon flux rate in excess of $P$ (units of photon cm$^{-2}$ s$^{-1}$) were discussed by Mészáros & Mészáros , 1995 (MM95) for arbitrary Friedman models with zero cosmological constant (in that paper $C$ was used for the photon flux, but here instead we use $P$ for the photon flux to avoid confusion with the more common usage of $C$ [ s$^{-1}$] as the count rate). As discussed in MM95, effects of a non-flat cosmology ($\Omega_o < 1$) are small, and to a first approximation can be neglected. Below we assume $\Omega_o = 1$ everywhere. The effect of a pure density evolution is approximated through a dependence

$$n(z) = n_o(1+z)^D ,  \qquad (1)$$

where $n$ is the physical burst density rate in cm$^{-3}$ yr$^{-1}$, $n_o$ is the density rate at $z=0$ ($D=3$ corresponds therefore to a non-evolving, constant comoving density). For a source emitting $\mathcal{L}$ photons s$^{-1}$ with a power law photon number spectrum $\mathcal{L}_\nu \propto \nu^{(\alpha-2)}$ (i.e. $\alpha=0$ corresponds to a flat power-per-decade spectrum), assuming most of the photons are collected in an energy range where $\alpha \sim$ constant a K-correction is necessary (e.g. Mao & Paczyński , 1993, Dermer, 1992). This can be folded in with the density evolution by using an effective scale factor exponent $D_{eff} = (D+\alpha-1)$ (MM95) and this K-correction is small or does not apply to most bursts, for which $\alpha \sim 1$ in the range 50-300 keV where BATSE collects most of the GRB photons used to determine the peak flux $P$ (e.g. Band, *et al.* , 1993). The photon luminosity function in the 50-300 keV range is represented by either one of the two forms,

$$\Phi(\mathcal{L}) = \begin{cases} n_o \delta(\mathcal{L} - \mathcal{L}_o) , & \text{(standard candle)} ; \\ \bar{n}\mathcal{L}_{min}^{-1}(\mathcal{L}/\mathcal{L}_{min})^{-\beta} & \text{for } \mathcal{L}_{min} \leq \mathcal{L} \leq \mathcal{L}_{max} \text{ (power law)} \end{cases} \qquad (2)$$

For the rest of the paper we take $\mathcal{L}$ to be the burst peak photon luminosity [ s$^{-1}$], $P$ to be peak photon flux [ cm$^{-2}$ s$^{-1}$], $n_o$ is the physical density of bursts per year at $z = 0$, $\bar{n} = n_o(1-\beta)/(K^{1-\beta} - 1)$ and $K = \mathcal{L}_{max}/\mathcal{L}_{min}$ gives the effective spread of the intrinsic luminosity function in the power law case.

The integral number distribution of bursts per year with peak flux rate above $P$ is given by (MM95) as

$$N(>P) = \frac{4\pi}{3} \frac{\mathcal{L}_e^{3/2} n_e}{(4\pi P)^{3/2}} \, I, \qquad (3)$$

where $\mathcal{L}_e = \mathcal{L}_o$, $n_e = n_o$ for the standard candle, $\mathcal{L}_e = \mathcal{L}_{min}$, $n_e = \bar{n}$ for the power law luminosity function, and $I$ is a dimensionless analytical function of $S$ and the model parameters, i.e. the luminosity function parameters $\mathcal{L}_o$ or $K, \mathcal{L}_{min}, \beta$, and the density evolution parameter $D$. In the redshift limited case, the maximum source redshift $z_{max}$ is an additional parameter, and the expression corresponding to equation (3) is given in the appendix of MM95. The differential number distribution $N(P) = -dN(>P)/dP$ can be obtained from the integral expressions through differentiation.

## 3. STATISTICAL MODEL FITS

For the numerical fits we used the BATSE 2B catalog available electronically. The 1024 ms peak fluxes $P$ (photons cm$^{-2}$ s$^{-1}$) were used, and only events with peak count rates divided by threshold count rates $C_{max}/C_{min} > 1$ were included, where $C_{min}$ is the published count threshold for each event. The 2B sample with this criterion consists of 278 entries in the catalog. Applying the efficiency tables published with the catalog to correct for detector inefficiency near the trigger threshold, the nominal number of bursts accumulated by BATSE over a period of two years with peak fluxes above $\log P \geq -0.6$ is 369. We chose for these bursts a binning equidistant in $\log_{10} P$, with step size 0.2 between -0.6 and 1.2, which gives 9 equal bins with a minimum number of 7 events per bin (in the highest $S$ bin, $\log_{10} P = 1.0$ to 1.2) for the two year 2B sample. The fits were made to the differential burst number distribution $N(P)$ as a function of peak photon flux $P$ (since only in the differential distribution may the bins be considered independent of each other for a $\chi^2$ fit) and the errors in each bin were taken to be the square root of the number of events in that bin.

Some of the fits were made using an extended 2B plus PVO sample. For the PVO events, we used the PVO portion of Table 2 of Fenimore and Bloom (1995) [FB95] for $\log_{10} P \geq 1.2$. A number of subtle issues concerning a matching between the different PVO and BATSE data sets are discussed by Fenimore, *et al.*, 1993, who indicate that systematic uncertainties of $\pm 10\%$ in the relative normalization cannot be ruled out. The matching of the level of the BATSE and PVO curves was taken directly from FB95. The PVO data was rebinned, ignoring PVO bursts below $\log_{10} P = 1.2$ so as not to count twice, and its level was renormalized so that the matching 2B data had the same level as in the original 2B catalog, i.e about 2 years. The errors for the PVO sample were also renormalized taking into account the fact that data had accumulated over more than ten years in the PVO case, keeping the relative errors the same. We used 5 bins in the PVO range, so that the combined 2B+PVO fits have 9+5=14 bins, reaching up to $\log_{10} P = 3.0$.

i) The SC fits (standard candle with density evolution) involve the fewest parameters: the photon luminosity $\mathcal{L}_o$ (ph/s), the density $n_o$ and the density evolution parameter $D$, under the brightness-limited assumption. For the 2B sample between peak fluxes $-0.6 \leq \log P \leq 1.2$ the free parameters are $p = 3$, the degrees of freedom are $f = 6$ and the best $\chi^2_{red}$ (reduced chi-square or $\chi^2$ divided by degrees of freedom) is 0.85 at the innermost mark. The $1\sigma, 2\sigma, 3\sigma$ significance contours were determined using the standard prescription (e.g., Press, *et al.*, 1986, or Lampton, Bowyer and Margon, 1976). The fit (Figure 1a) is good over an elongated region describing a relation between the luminosity and the density evolution. For faster density evolution $n \propto (1 + z)^D$ (larger $D$) the luminosity must increase because most sources are at larger redshifts, while for slower or negative evolution the luminosity must decrease, since most sources are at small redshift ($D = 3$ is constant comoving density). The optimal fit is obtained for $D = 3.5$ and $\mathcal{L}_o \sim 10^{57}$ s$^{-1}$. This luminosity is close to the SC value deduced, e.g. by MM95 and some previous authors as well (corresponding to $L_o \sim 10^{51}$ erg s$^{-1}$ for typical photon energies of 0.5 MeV). However the preference for $D = 3.5$ was not, as far as we can tell, encountered in previous fits. The $\chi^2_{red}$ around the best fit minimum is 0.85; however, the $1\sigma$ region around it is rather large, even if not very wide, so this preference is not strong.

For the SC fits using the 2B+PVO sample, $p = 3$, $f = 11$ and the fits are shown in Fig. 1b, with a best $\chi^2_{red} = 0.62$ at the central mark enclosed by its $1\sigma, 2\sigma, 3\sigma$ contours. We note that this ignores any possible systematic errors in matching BATSE and PVO beyond what is done in Fenimore *et al.*, 1993, and Fenimore and Bloom, 1995. If any extra errors were present, they could in principle increase the size of the significance regions discussed below (e.g. it might add an extra free parameter for the relative normalization). However, such errors are extremely difficult to quantify without going into additional details of the instruments, and we follow Fenimore and Bloom (1995) in adopting their relative normalization as adequate without further manipulation. The effect of the rare high flux PVO bursts satisfying a tight $N \propto P^{-3/2}$ correlation at $1.2 \lesssim \log_{10} P \lesssim 3.0$ is to improve the best fit (lower $\chi^2_{red}$) and to place it at a somewhat smaller luminosity $\mathcal{L}_o \sim 5 \times 10^{56}$ s$^{-1}$ and closer to comoving constant density evolution, $D \sim 3$. This is in good agreement with Fenimore and Bloom's (1995) value of $L_o \sim 5 \times 10^{50}$ erg s$^{-1}$. However, as seen from Fig. 1b, the $1\sigma$ region around this best fit minimum is compatible with both larger and smaller $\mathcal{L}_o$ and $D$. In contrast to the pure 2B fit, however, the joint $1\sigma$ upper limit for $\mathcal{L}_o$ and $D$ are $\mathcal{L}_o \lesssim 5 \times 10^{57}$ s$^{-1}$, $D \lesssim 4.5$ (or $3\sigma$ joint upper limits $\mathcal{L}_o \lesssim 5 \times 10^{58}$ s$^{-1}$, $D \lesssim 5$).

ii) The PL fits (Power-Law luminosity function bounded between $\mathcal{L}_{min}$ and $\mathcal{L}_{max}$ and including density evolution) are shown in Figures 2 and 3 in the brightness limited case. The parameters of the fits are $\mathcal{L}_{max}$, $K$, $n_o$, $D$, where $K = \mathcal{L}_{max}/\mathcal{L}_{min}$, and for the 2B sample $p = 4$, $f = 5$. If the index $\beta$ is taken as an additional variable parameter, the fits maximize at slopes significantly steeper than -5/2, and this results in $\mathcal{L}_{min}$ dominating the luminosity function over the whole range of $S$, giving essentially a standard candle case. As discussed in MM95 (see also Ulmer and Wijers, 1995, Wasserman, 1992), this is because a luminosity function slope $\beta = -5/2$ reproduces directly the Euclidean integral distribution slope -3/2 (which is a differential distribution of slope -5/2). However, the quality of the fits with such very steep slopes (or SC cases) is not significantly different from those for a fixed slope of $\beta \sim 1.88$, in the sense that in the brightness limited case both give $\chi^2_{red} < 1$. An index close to -1.88 is suggested for a power law luminosity slope, since the slope of the integral distribution at low $S$ is approximately -0.88 (e.g. Meegan *et al.*, 1992; see also Wasserman, 1992). An illustration of this is given, e.g. in Fig. 1 of MM95. Either steeper or shallower slopes would lead to an effective standard candle case below $P_{turn} \sim 6$ cm$^{-2}$ s$^{-1}$. Since there is no way to either rule out or prefer the physically interesting slope of -1.88 in the brightness limited case, we discuss here fits which take a fixed value $\beta = 15/8 \sim 1.88$. Given such an index the interesting question is what can be said about the lower and upper limits $\mathcal{L}_{min}$ or $\mathcal{L}_{max}$, or $\mathcal{L}_{max}$ and the intrinsic spread $K$, and how do the fits compare with SC fits or $K = 1$ fits of the same slope. The 2B PL fits (Figure 2) in the $\mathcal{L}_{max}, K$ parameter plane show a broad inverse correlation between the allowed values of $\mathcal{L}_{max}$ and $K$. For increasing values of the density evolution index $D$ the required values of $\mathcal{L}_{max}$ also increase, as one would infer from the previous SC fits, since the sources are more distant and need to be more luminous. The best $\chi^2_{red}$ for the cases $D = 2, 3, 4$ (in panels 2a,2b,2c) are $\sim 0.82$ in all three cases, at the (varying) location of the innermost mark surrounded by the $1\sigma, 2\sigma, 3\sigma$ contours. Fits have also be obtained for other values of $-2 \leq D \leq 5$, the trend being apparent from the three values shown here. These 2B best fits are obtained for values of $\mathcal{L}_{max}, K \sim 2 \times 10^{57}, 30$ for $D = 2$, $\mathcal{L}_{max}, K \sim 3 \times 10^{57}, 15$ for $D = 3$, $\mathcal{L}_{max}, K \sim 8 \times 10^{57}, 5$ for $D = 3$. However, the 2B fits are not very strongly constrained in the Euclidean -3/2 region, and the joint $1\sigma$ limits could extend to relatively large values of $\mathcal{L}_{max}$ and of the intrinsic luminosity ratio $K$.

The similar PL fits for the 2B+PVO sample in the brightness-limited case, with $p = 4$, $f = 10$, are shown in Figs. 3a,b,c for $D = 2, 3, 4$. The presence of the bright PVO bursts constrains the high luminosity portion of the fits, and brings the best fit minima towards somewhat lower luminosities $\mathcal{L}_{max}$ and lower luminosity ratios $K$. The best fit joint minima are $\chi^2_{red} = 0.62, 0.62, 0.70$ near $\mathcal{L}_{max}, K \sim 7 \times 10^{56}, 10$ for $D = 2$, $\mathcal{L}_{max}, K \sim 8 \times 10^{56}, 3$ for $D = 3$, $\mathcal{L}_{max}, K \sim 1.5 \times 10^{57}, 1$ for $D = 4$. While the best fit values of $K$ are relatively small, it is to be noted that for all three $D$ the joint $1\sigma$ upper limits allow intrinsic luminosity ratios of order $K \sim 10^2$, while the $3\sigma$ upper limits are of order $K \sim 10^3$.

iii) The PLZ fits (luminosity function with density evolution and redshift cutoff model) differ from the above in that a maximum redshift $z_{max}$ is included as a parameter in the expression for the distribution N (in the integral I of equation [3], see MM95). The parameters are $\mathcal{L}_{max}, K, n_o, D, z_{max}$, so $p = 5$, and for the BATSE 2B sample ($f = 4$) the fits are shown in Figs. 4a through 4f, for D=2,3,4 and two particular choices of $z_{max} = 2, 6$. The best $\chi^2_{red}$ values marked inside the $1\sigma, 2\sigma, 3\sigma$ contours are 1.1, 1.0, 1.2 for $D = 2, 3, 4$ ($z_{max} = 2$) and 1.2, 1.2, 0.8 for $D = 2, 3, 4$ ($z_{max} = 6$). In the $\mathcal{L}_{max}, K$ plane the results are shown in Figure 4a,b,c (left panels) for the case $z_{max} = 2.0$, while Figure 4d,e,f (right panels) is for $z_{max} = 6.0$. Three values of the evolution index $D$ are shown. In the $D = 2$ case most of the sources are nearby, so the 2B fits with $z_{max} = 2$ (Figure 4a) are not very different from those for $z_{max} = \infty$ (Figure 2a) except for somewhat lower $\mathcal{L}_{max}, K$ values. For $D = 4$ however, most sources would be farther and the $z_{max} = 2$ case (Figure 4c) restricts the low luminosities and requires higher $\mathcal{L}_{max}, K$ values than the brightness-limited $z_{max} = \infty$ case of Figure 2c. For $z_{max} = 6$, however, the results (Figure 4 d,e,f) are fairly close to the brightness limited case (2a,b,c), except for $D = 4$ where there is some restriction at low luminosities.

For the 2B+PVO sample, the PLZ power-law maximum redshift fits provide some additional restrictions in the $z_{max} = 2$ case. The PLZ fits are shown in Figs. 5a through f, where $p = 5$, $f = 9$, and the $\chi^2_{red}$ values are 0.7, 0.7, 0.9 for $z_{max} = 2$, and 0.7, 0.8, 0.7 for $z_{max} = 6$. While $D = 2$ is fairly similar for both redshifts (sources are close by in both due to evolution) the $D = 4$ case (sources far due to evolution) is constrained by the PVO data to have larger $K$ and $\mathcal{L}_{max}$ values, and gives a joint $3\sigma$ lower limit of $K \gtrsim 5$ for the $z_{max} = 2$ case. For the $z_{max} = 6$ case, however, the 2B+PVO fits (Figure 5d,e,f) are not very different from those in the brightness limited case (Figure 3), except for a not too significant preference towards somewhat larger $\mathcal{L}_{max}, K$.

# 4. DISCUSSION

The fits presented above show that a cosmological interpretation is compatible with the data under a variety of assumptions. Good fits to the observed differential distribution of bursts $N(P)$ as a function of peak photon flux $P$ are obtained both under a standard candle (SC) and under a power-law (PL) luminosity function assumption, as well as for redshift-limited power-law luminosity function (PLZ). Fits were obtained for a range of density evolution indices $D$, defined through a physical density dependence $n_o \propto (1+z)^D$ where $D = 3$ is equivalent to a non-evolving, constant comoving density case (i.e. $\delta = D - 3 = 0$, where $\delta$ is the index for the *comoving* density $n_{com} = n_0(1+z)^\delta$ sometimes used in the literature). The BATSE 2B fits, which have many weak bursts but few very bright bursts, do not constrain $D$ except through an inverse functional dependence on the luminosity. For the 2B+PVO fits, which include a number of PVO bright bursts, a $3\sigma$ upper limit to the SC luminosity and the evolution index are obtained, $\mathcal{L}_o \lesssim 5 \times 10^{58}$ s$^{-1}$, $D \leq 5$, with optimal values at $\mathcal{L}_o \sim 5 \times 10^{56}$ s$^{-1}$, $D \sim 3$.

In the 2B and 2B+PVO samples there is no conclusive evidence for a lower cutoff $P_{min}$ (or a flattening of the integral distribution), the uncertainty being due to large and uncertain trigger corrections near threshold. In the brightness-limited case, the lowest peak flux $P$ used also corresponds to the largest redshift observed, via the relation $\mathcal{L}_o = 4\pi R_o^2[(1+z)^{1/2} - 1]^2 P = 4 \times 10^{57} h^{-2}[(1+z)^{1/2} - 1]^2 P$ s$^{-1}$, where $R_o = 2c/H_o$ is the Hubble radius, and $H_o = 100~h$ km/ s/ Mpc is the present Hubble constant. Thus by re-plotting Figure 1 as a function of $z(P_{min})$ and taking into account differences in notation for $D$ one can also obtain a plot similar to the figure 3 of Cohen and Piran (1995), although the sample criteria and cuts are somewhat different. Our best fit $\mathcal{L}_o$ varies with $D$ (Figures 1a,b), and $\log_{10} P_{min} = -0.6$ corresponds in Figure 1b to $z_{max} \gtrsim 1.9$ (10.; 0.8) for our best fit ($\pm 1\sigma$) values of $\mathcal{L}_o \sim 5 \times 10^{56}$ s$^{-1}$, $D \sim 3$ ($\mathcal{L}_o \sim 5 \times 10^{57}$ s$^{-1}$, $D \sim 4.5$; $\mathcal{L}_o \sim 1 \times 10^{56}$ s$^{-1}$, $D \sim 0.0$). (To be specific, here we arbitrarily took a lower value of $D \sim 0$ but from Fig. 1b one sees that the lower limit could actually be smaller and could extend to $D \lesssim -2$ and values of $\mathcal{L}_o \lesssim 10^{56}$ s$^{-1}$). Thus, if an intrinsic energy-stretching of the time profiles exists indicating a maximum SC redshift $z_{max} \sim 6$ (as argued by Fenimore and Bloom, 1995), this could be easily accommodated within our 2B+PVO $1\sigma$ SC limits with a density evolution faster than comoving constant, $D \geq 3$.

The fits with a power law (PL) luminosity function (equation [2]) bounded between $\mathcal{L}_{min}$ and $\mathcal{L}_{max}$ behave in a manner which is qualitatively similar to that in the SC case. The fits presented are for a luminosity function index $\beta = 15/8$. In the brightness limited case using the BATSE 2B sample, there is an inverse correlation between maximum luminosity and the evolution index $D$. (As discussed in Mészáros and Mészáros, 1995 [MM95], for the Euclidean $P^{-3/2}$ part of the integral distribution the behavior is dominated by the large luminosity sources $\mathcal{L} \sim \mathcal{L}_{max}$, if $\beta < 5/2$). While for 2B there is a preference for a ratio of intrinsic luminosities $K = \mathcal{L}_{max}/\mathcal{L}_{min}$ of order 10-30, the $1\sigma$ upper limits are compatible with much higher values. However, using the PVO data as well, we obtain more specific constraints on $K$ and $\mathcal{L}_{max}$. For $D = 2, 3, 4$ the best fit $K$ and its $1\sigma$ upper limits are $(5, 100)$; $(2, 60)$; $(1, 30)$ (see Figure 3). These results for $D = 3$ (nonevolving density) are in significant agreement with the results obtained from the method of moments on the observed luminosity distribution by Emslie & Horack (1994), Horack, Emslie & Meegan (1995) and Horack, Emslie and Hartmann (1995), using the 2B sample. They are also compatible with the Euclidean distribution results of Hakkila, *et al.* (1994) using 2B, as well as results on the observed and/or intrinsic luminosity distributions by Ulmer & Wijers (1995) using 2B, Ulmer, Wijers & Fenimore (1995) using 2B+PVO data and Hakkila, *et al.* (1995) using 3B + PVO data. It is worth noting that while some of these groups have put limits of a factor $\sim 10$ for the width containing 90% of the bursts in the *observed* luminosity distribution, our widths here refer to the *intrinsic* luminosity distribution. Also, note that our width limits apply for the particular luminosity function slope $\beta \sim 15/8$ which fits the low flux end of the observed number distribution. As emphasized in MM95, for such a slope each luminosity decade contains only about 10% as many bursts as the previous lower luminosity decade, so $\sim 90\%$ of the bursts are automatically in the lowest decade. For other slopes $\beta < 1$ or $> 5/2$ the intrinsic widths could be much larger, since the upper or lower ends of the luminosity function dominate and one is dealing effectively with standard candles.

By assuming a redshift cutoff to the source distribution, the corresponding power-law luminosity function fits (PLZ) involve more than just a simple relation relation between $P_{min}$ and $z_{max}$. As discussed in MM95, even for $P > P_{min}$ the integrations over the luminosity functions depend on $z_{max}$ in a nontrivial manner. Redshift-limited power law fits were carried out for a variety of redshifts, of which two particular values $z_{max} = 2.0, 6.0$ are shown in figures 4 (2B) and 5 (2B+PVO) for three values of the evolution index $D$. The case $z_{max} = 2$ is representative of the redshift inferred by Norris, *et al.* (1995) based on an analysis of BATSE 2B time profiles and brightnesses under the assumption of cosmological time-dilation and redshift, without allowance for any possible intrinsic energy-stretching of the profiles. The values for $z_{max} = 6$ are characteristic of the maximum redshift inferred by Fenimore and Bloom (1995) if there is such an intrinsic energy stretching. We note that Mitrofanov, *et al.* (1994) have found no evidence for a cosmological time dilation, while Norris, *et al.* (1995) find no strong need for intrinsic energy stretching (see however also Fenimore, *et al.* , 1995). The 2B fits are not bounded from above in the $K, \mathcal{L}_{max}$ plane, due to the lack of very strong bursts in this sample, but especially for the cases of strong evolution ($D > 3$, sources preferentially distant) there is a lower bound to the maximum luminosity $\mathcal{L}_{max}$ which is particularly strong for low $z_{max}$, e.g. for $z_{max} = 2$, $D = 4$ the $3\sigma$ limit is $\mathcal{L}_{max} \lesssim 10^{57}$ s$^{-1}$ (figure 4c). With the 2B+PVO sample, the PLZ fits are constrained from above ($K, \mathcal{L}_{max}$ plane) in all cases considered, as seen in figure 5. This is because the strong sources from PVO follow a well defined -3/2 integral distribution behavior. For both $z_{max} = 2$ and 6, the $3\sigma$ upper limits of $K, \mathcal{L}_{max}$ are $\sim 10^3$, $1 - 3 \times 10^{58}$ s$^{-1}$ depending on $D$ (figure 5). For $D \lesssim 2$ or for $z_{max} \gtrsim 2$, there is no restriction against $K \sim 1$ (i.e. standard candle) or against $\mathcal{L}_{max} \sim \mathcal{L}_o \lesssim 10^{57}$ s$^{-1}$. However, the 2B+PVO fits constrain $K, \mathcal{L}_{max}$ from below for $D \gtrsim 3$ and $z_{max} \lesssim 2$ (figure 5b,c). For $D = 3$ we find $K \gtrsim 1$, $\mathcal{L}_{max} \gtrsim 10^{57}$ s$^{-1}$, while for $D = 4$ we find $K \gtrsim 3$, $\mathcal{L}_{max} \gtrsim 2 \times 10^{57}$ s$^{-1}$ as the $3\sigma$ lower limits.

A truncated power law is a highly idealized luminosity function. Nonetheless, it is a form commonly found in astrophysics, and its spread can give us some idea about how standard is the efficiency of the source in producing $\gamma$-rays. While it is easy to envisage a "standard" energy, e.g. $\sim M_\odot c^2$, it would be more difficult to envisage converting that into $\gamma$-rays with a well defined efficiency which is the same in every source. It is thus reassuring that, while previous fits have found standard candle models acceptable, and limits of a factor $\sim 10$ can be put on the spread of the distribution containing 90% the *observed* low flux bursts, equally good fits are found for a significant spread ($K \gtrsim 10$ and up to 300 or $10^3$ at the $1\sigma$ or $3\sigma$ level) in the *intrinsic* luminosity distribution with a slope matching the low flux number distribution. A lower limit on the spread would also be interesting, since it might say something about the possible range of masses involved. On the other hand, if the luminosity spread was constrained to be large, it might be surprising that the turnover of the counts at low $P$ below the $-3/2$ behavior is not even more gradual than what is observed. However, in most cases there is no need for the spread to be very large. We find a $3\sigma$ lower limit $K \gtrsim 3$ only for $z_{max} \lesssim 2$ and $D \gtrsim 4$, but for $z_{max} \gtrsim 2$ and $D \lesssim 4$ there is no significant difference to within $3\sigma$ between values of $1 \lesssim K \lesssim 10^3$. Such conclusions, of course, depend on the accuracy and completeness of the data at low $P$, and one cannot rule out that there may be some as yet undetermined corrections affecting the low $P$ counts. A discussion of incompleteness issues and methodological questions is given in Loredo and Wasserman, 1995. Additional data may affect any conclusions reached here. Here we have confined ourselves to the use of the published 2B data and correction tables.

The cosmological fits obtained are of good quality ($\chi^2_{red} \lesssim 1$) for a range of plausible model assumptions, including both standard candle and truncated power law luminosity functions in the brightness or redshift limited cases. The present analysis provides a discussion of the statistical significance of such fits including specific density evolution and luminosity function parameterizations. The constraints obtained here on the evolution parameter and the intrinsic luminosity function spread are stronger than in previous analyses due to the inclusion of both BATSE and PVO information, and the use of standard deviation measures. However, such constraints do not impose a fine-tuning problem, as the allowed parameter space region is of moderate size and includes a substantial range of physically reasonable values under the cosmological interpretation.

*Acknowledgements:* This research was supported through NASA NAG5-2857, NAG5-2362, and (I.H.) Soros Foundation and Eötvös Foundation fellowships. We are grateful to H. Papathanassiou and V. Karas for expertise, and to C. Meegan, E. Feigelson, J. Nousek and M.J. Rees for useful advice.

## Figure Captions

*Figure 1* : Standard Candle (SC) cosmological fits for $\Omega_o = 1$, $\Lambda = 0$ for photon luminosity $\mathcal{L}_o$ (photon/s) and physical density evolution $n(z) \propto (1+z)^D$ (where $D = 3$ is the nonevolving, comoving constant density case). The inner dots are the best fit $\chi^2_{red}$ minimum location, with $1\sigma, 2\sigma, 3\sigma$ contours increasing outwards. a) Top: using the BATSE 2B data base; b) Bottom: using the 2B plus PVO information (see text).

*Figure 2* : Power law (PL) luminosity function cosmological fits for $\Omega_o = 1$, $\Lambda = 0$, luminosity index $\beta = 15/8$ bounded between $\mathcal{L}_{min}$ and $\mathcal{L}_{max}$ and physical density evolution $n(z) \propto (1+z)^D$ n the brightness limited case, using the 2B sample. The abcissa is $\log_{10} \mathcal{L}_{max}$ and the ordinate is $\log_{10} K = \log_{10}(\mathcal{L}_{max}/\mathcal{L}_{min})$. The three panels a,b,c from top to bottom are for $D = 2, 3, 4$.

*Figure 3* : Power law (PL) luminosity function fits, same parameters as for figure 2, brightness limited case, but for the 2B+PVO sample.

*Figure 4* : Power law luminosity function (PLZ) fits to the 2B sample for the redshift limited case (sources with density evolution and luminosity function as fig. 3, $\beta = 15/8$, but only out to finite $z_{max}$). Left panels: $z_{max} = 2.0$, from top to bottom a),b),c) cases $D = 2, 3, 4$; right panels: $z_{max} = 6.0$, from top to bottom c),d),e) cases $D = 2, 3, 4$.

*Figure 5* : Power law luminosity (PLZ) fits to the 2B+PVO sample, $\beta = 15/8$, for the redshift limited case (other details as for figure 4). Left panels: $z_{max} = 2.0$, from top to bottom a),b),c) cases $D = 2, 3, 4$; right panels: $z_{max} = 6.0$, from top to bottom c),d),e) cases $D = 2, 3, 4$.

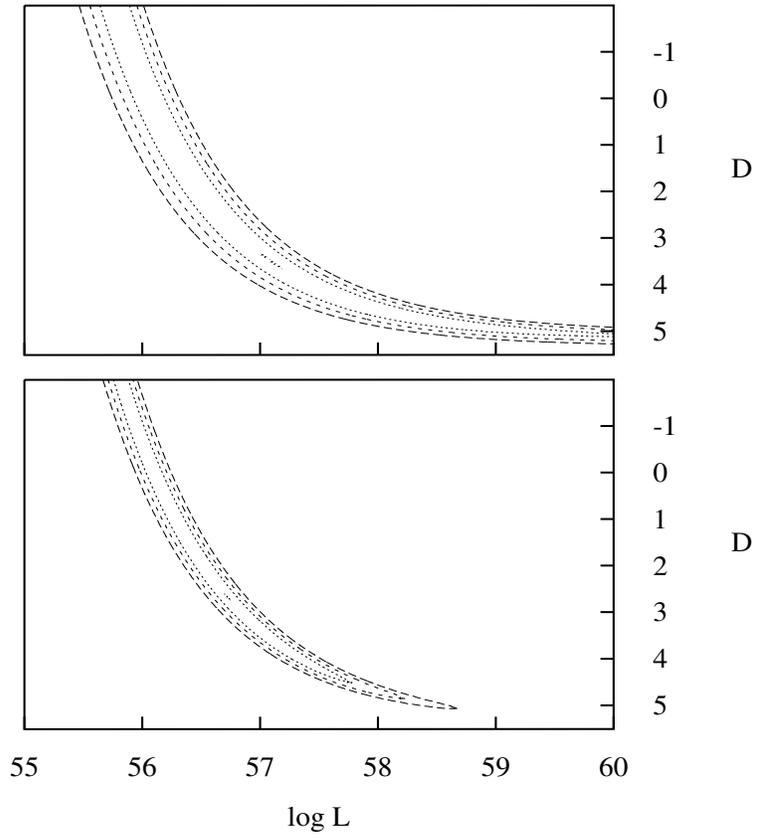

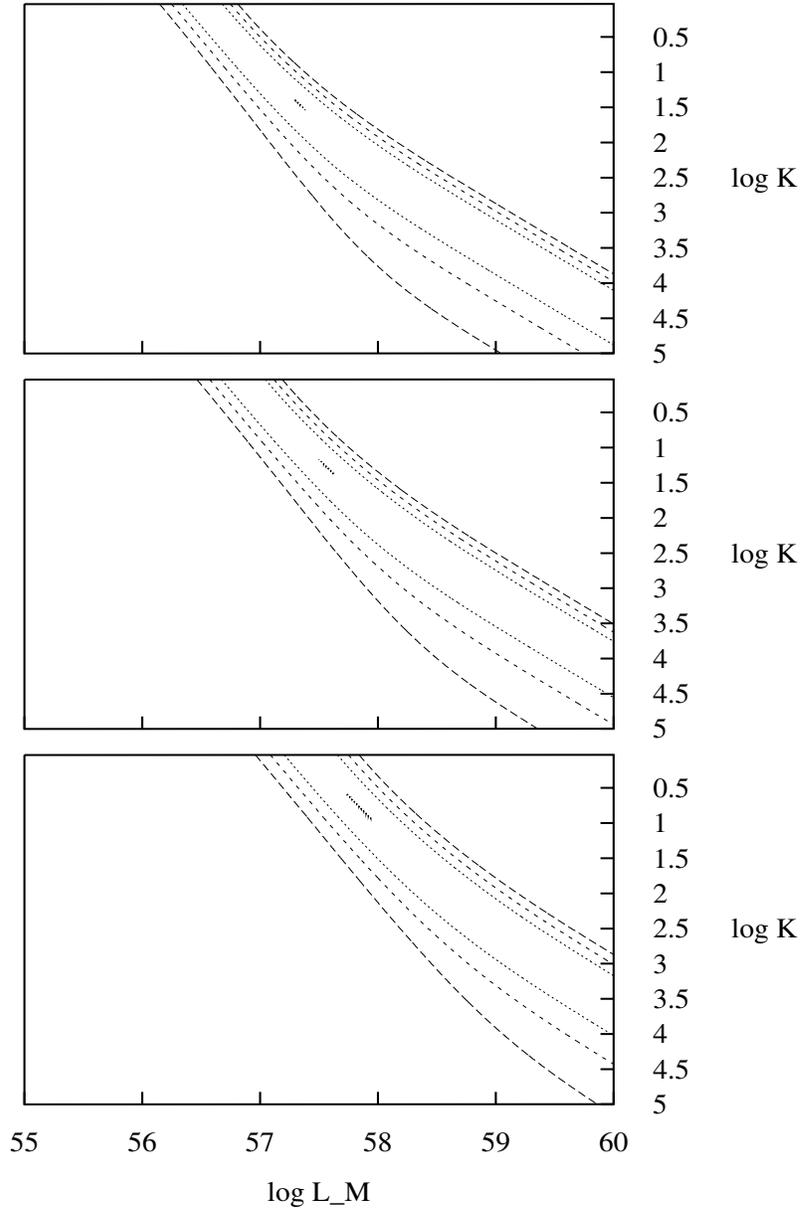

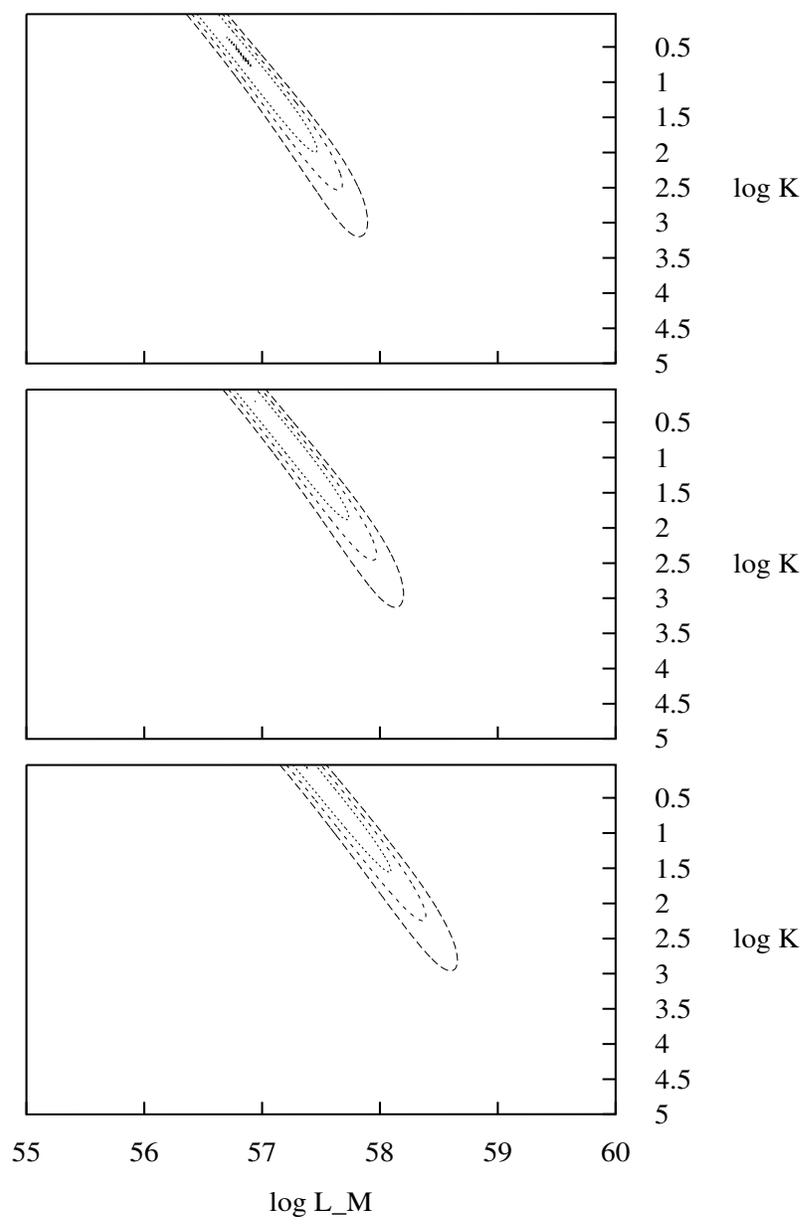

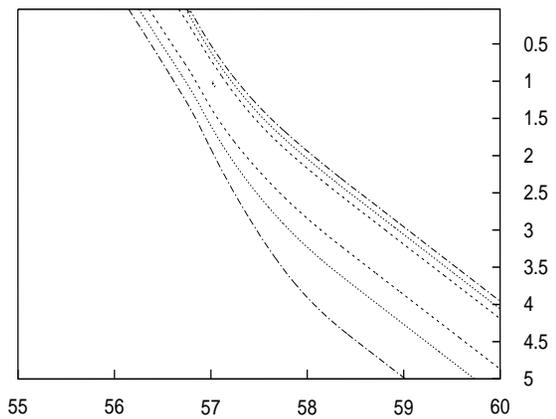
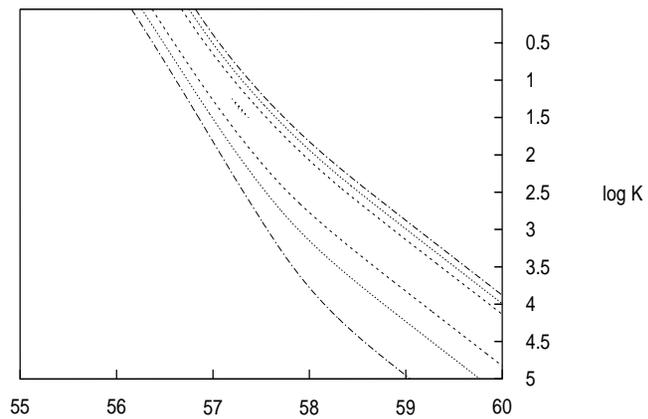
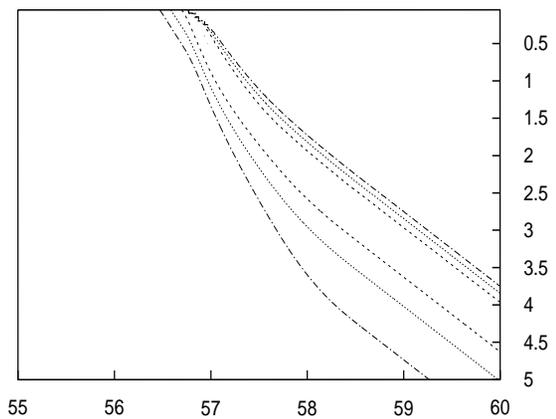
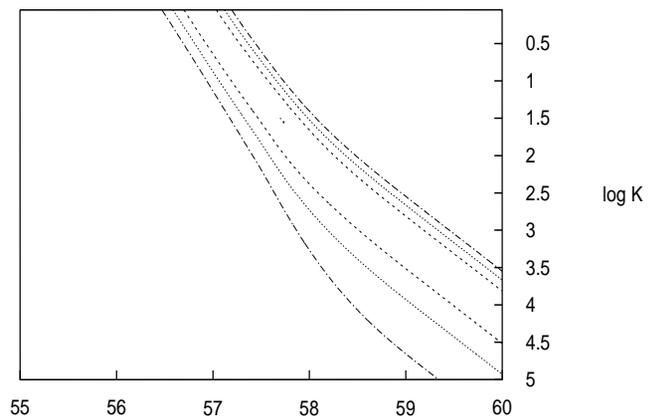
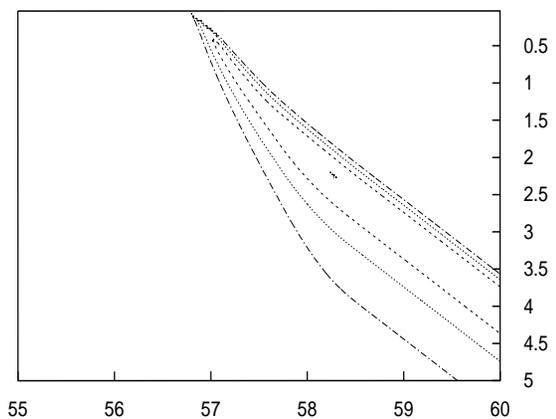
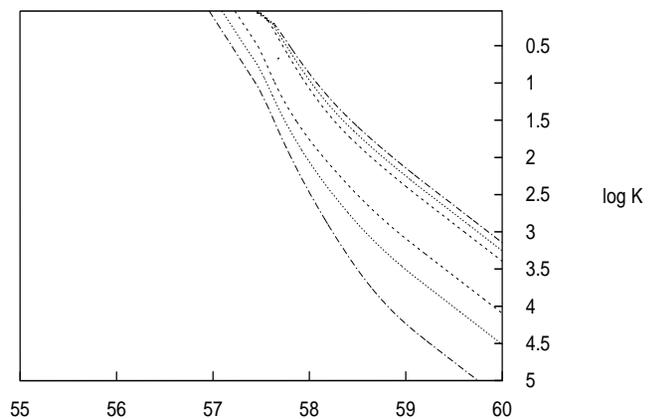

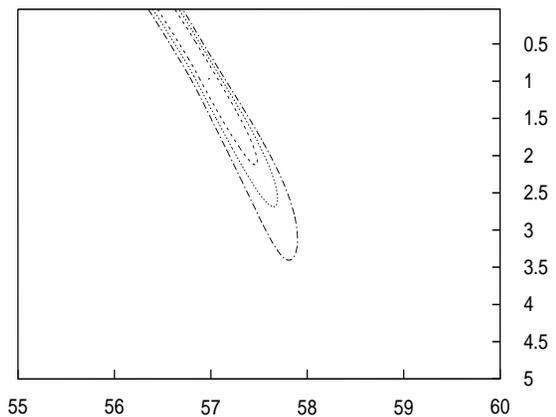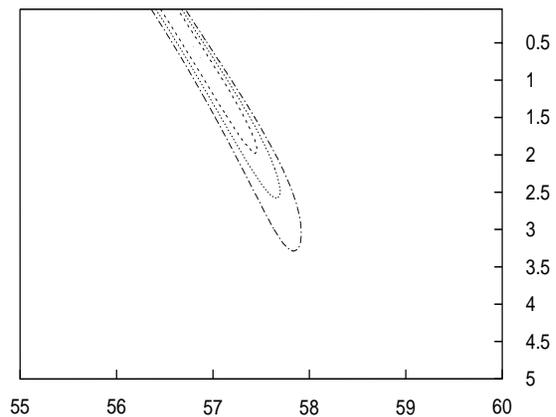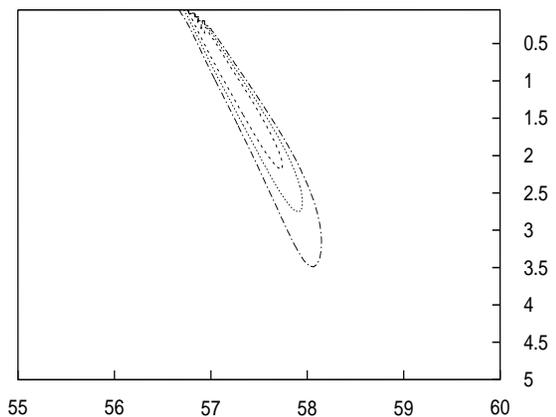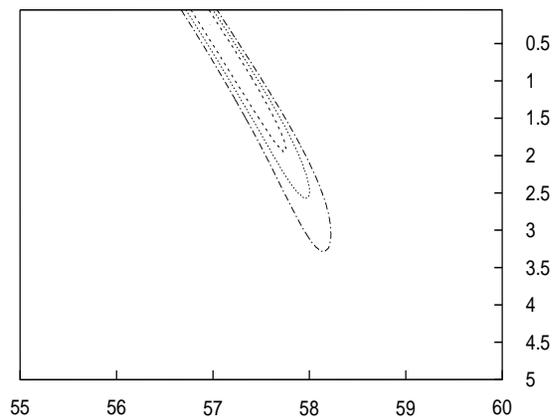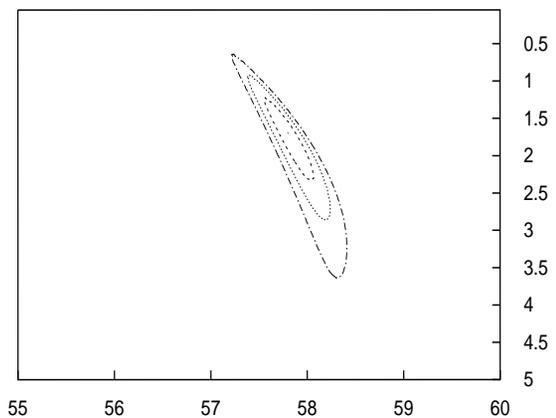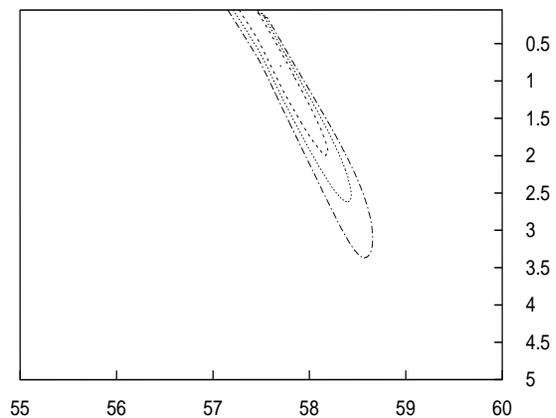